\documentclass[]{aipproc}
\usepackage[T1]{fontenc}
\usepackage[latin9]{inputenc}
\usepackage{units}
\usepackage{amsmath}
\usepackage{graphicx}
\usepackage[english]{babel}

\makeatletter

\newcommand{\lyxdot}{.}

\layoutstyle{8x11single}

\makeatother

\usepackage{babel}
\begin{document}
\title[Wet separation of particles]{Numerical analysis of wet separation of particles by density differences}
\author{D.~Markauskas}{
  address={Ruhr University Bochum, Universitaetsstrasse 150, D-44780 Bochum, Germany},
  email={markauskas@leat.rub.de},
}
\author{H.~Kruggel-Emden}{
  address={Ruhr University Bochum, Universitaetsstrasse 150, D-44780 Bochum, Germany},
  email={kruggel-emden@leat.rub.de},
}

\begin{abstract}
Wet particle separation is widely used in mineral processing and plastic recycling to separate mixtures of particulate materials into further usable fractions due to density differences. This work presents efforts aiming to numerically analyze the wet separation of particles with different densities. In the current study the discrete element method (DEM) is used for the solid phase while the smoothed particle hydrodynamics (SPH)  is used for modeling of the liquid phase. The two phases are coupled by the use of a volume averaging technique. In the current study, simulations of spherical particle separation were performed. In these simulations, a set of generated particles with two different densities is dropped into a rectangular container filled with liquid. The results of simulations with two different mixtures of particles demonstrated how separation depends on the densities of particles.
\end{abstract}

\date{\today}

\maketitle

\section{Introduction}

Wet particle separation is widely used in mineral processing and plastic
recycling to separate mixtures of particulate materials into further
usable fractions due to density differences. Despite its wide usage,
the wet particle separation process is often attributed to operational
problems. Difficulties arise in float-sink separation if density differences
between plastic fractions become low \cite{Menad2013} or in case
of elevated feed rates. A review of varying separation technologies
and their efficiencies can be found in \cite{Dodbiba2004}. However,
a review of the state of the art indicates that numerical modeling
has not yet been applied to wet separation processes due to the lack
of applicable numerical schemes. The current research is aimed to
fill this gap by developing the numerical tool and its application
to wet separation of particles.

A smoothed particle hydrodynamics (SPH) method is used for modeling
of the liquid phase. The principal idea of SPH is to treat the fluid
in a completely mesh-free fashion, in terms of a set of sampling particles
\cite{Monaghan2005}. SPH particles represent a finite mass of the
discretized fluid and carry all information about physical variables
evaluated at their positions. Function values at a fluid particle
are interpolated from function values at surrounding particles using
a kernel function and its derivative. Because of the mesh-free nature
of SPH, it can be used for dealing with problems where large displacements
of the fluid-structure interface and a rapidly moving fluid free-surface
are present \cite{Prakash2015,Sivanesapillai2014}.

The discrete element method (DEM) is used for modeling of solid particles.
The DEM is a Lagrangian method introduced by Cundall and Strack \cite{Cundall1979}
to describe granular materials and is nowadays widely used for modeling
particulate flows \cite{Zhu2008}. In this method, contact forces
are calculated for particles interacting with their nearest neighbors
and with the walls based on appropriate contact laws while the motion
of the particles follows Newton's second law. The interaction of the
particles can be modeled using simple spring contact models, or using
Hertz theory for modeling the normal interactions and the theory of
Mindlin and Deresiewicz to model tangential interaction \cite{Kruggel-Emden2007,Kruggel-Emden2008}.
Different particle shapes can be used in the DEM, however spherical
particles dominate because these particles are easy to describe by
their center of mass and radius and their use is computationally very
efficient due to straight forward contact detection. 

The volume-averaging technique is used for the coupling of the SPH
with the DEM. This technique based on the locally averaged Navier-Stokes
equations was first reported by Tsuji et al. \cite{Tsuji1993} in
their work where a finite volume method was coupled with the DEM.
A two-way coupling scheme between the DEM and the SPH has been derived
by Sun et al. \cite{Sun2013} and Robinson et al. \cite{Robinson2014}
which is applied here. 

In the current study wet separation of particles is simulated numerically.
The used numerical methodology, in which the DEM is used for the solid
phase and the SPH is used for the liquid phase, is described in the
following section. In the later section the numerical modeling is
presented, where the separation of spherical particles with different
densities is analyzed.

\section{Numerical methodology\label{sec:Numerical_methodol}}

For modeling of the fluid the continuity equation and the momentum
equation are used:

\begin{equation}
\frac{d\bar{\rho_{f}}}{dt}+\nabla\cdot(\bar{\rho}_{f}\mathbf{u}_{f})=0,\quad\frac{d\bar{\rho}_{f}\mathbf{u}_{f}}{dt}=-\nabla p+\nabla\cdot(\varepsilon\boldsymbol{\tau})-\mathbf{f}^{int}+\bar{\rho}_{f}\mathbf{g}\textrm{,}\label{eq:momen-continuity-Lag}
\end{equation}
where $\overline{\rho}_{f}=\varepsilon\rho$ is the superficial fluid
density of the fluid, $\varepsilon$ is the local mean fluid volume
fraction.

In the SPH the fluid is represented by separate particles. These particles
carry variables such as velocity, pressure and mass. No connectivity
is modeled between particles. The integral representation of the function
is approximated by summing up the values of the neighboring particles
using smoothing kernel functions.

The weekly compressible formulation of the SPH is used to simulate
an incompressible fluid. In SPH the continuity equation and the momentum
conservation equation (\ref{eq:momen-continuity-Lag}) for the fluid
particle $a$ takes the form 

\begin{equation}
\frac{d\bar{\rho}_{a}}{dt}=\underset{b}{\sum}m_{b}\mathbf{u}_{ab}\cdot\nabla_{a}W_{ab},\quad\begin{array}{c}
\frac{d\mathbf{u}_{a}}{dt}=-\underset{b}{\sum}m_{b}\left(\frac{p_{a}}{\bar{\rho}_{a}^{2}}+\frac{p_{b}}{\bar{\rho}_{b}^{2}}\right)\nabla_{a}W_{ab}+\mathbf{g}+\underset{b}{\sum}m_{b}\frac{\nu(\bar{\rho}_{a}+\bar{\rho}_{b})}{\bar{\rho}_{a}\bar{\rho}_{b}}\cdot\frac{\mathbf{r}_{ab}\nabla_{a}W_{ab}}{|\mathbf{r}_{ab}|^{2}+\delta^{2}}\mathbf{u}_{ab}+\frac{\mathbf{f}_{a}^{int}}{m_{a}}.\end{array}\label{eq:cont-mom-SPH}
\end{equation}
where indexes $a$ and $b$ indicate fluid particles. $m$ is the
mass. $\mathbf{u}_{ab}=\mathbf{u}_{a}-\mathbf{u}_{b}$ is the relative
velocity between particles $a$ and $b$. $\nabla_{a}W_{ab}=\nabla_{a}W(r_{a}-r_{b},h)$
is the gradient of the kernel function. $r_{a}$ and $r_{b}$ are
positions of the particles $a$ and $b$. The summation is performed
over all neighboring particles $b$ of particle $a$. Here a viscous
term introduced by Morris et al. \cite{Morris1997} is used, where
$\nu$ is the kinematic viscosity. $\delta$ is a small number used
to keep the denominator non-zero.

$\mathbf{f}_{a}^{int}$ in Eq. (\ref{eq:cont-mom-SPH}) is the solid-fluid
interaction force acting on the fluid particle $a$ due to the surrounding
solid particles:

\begin{equation}
\mathbf{f}_{a}^{int}=\underset{i}{\sum}-\nicefrac{V_{a}W_{ai}}{(\underset{b}{\sum}V_{b}W_{bi})}\mathbf{F}_{i}^{int}\,,\label{eq:f_int_S-F}
\end{equation}
where $V_{a}$ and $V_{b}$ are the volumes of fluid particles $a$
and $b$, while $\mathbf{F}_{i}^{int}$ is the solid-fluid interaction
force acting on solid particle $i$. The fluid volume fraction $\varepsilon_{a}$
is calculated from the volumes of all solid particles $i$ which are
in the smoothing domain of the fluid particle:

\begin{equation}
\varepsilon_{a}=1-\sum_{i}V_{i}W_{ai}\:,\label{eq:fluid-fraction}
\end{equation}
where $V_{i}$ is the volume of the solid particle $i$, while $W_{ai}=W(r_{a}-r_{i,h})$
is the kernel function.

The solid phase is modeled using the discrete element method. The
motion of solid particles is described by Newton's second law:

\begin{equation}
m_{i}\frac{d^{2}\mathbf{r}_{i}}{dt^{2}}=\mathbf{F}_{i}^{c}+\mathbf{F}_{i}^{g}+\mathbf{F}_{i}^{int},\label{eq:sp_newton}
\end{equation}
where $\mathbf{r}_{i}$ is the position of the solid particle, $\mathbf{F}_{i}^{c}$
is the contact force, $\mathbf{F}_{i}^{g}$ is the gravity force and
$\mathbf{F}_{i}^{int}$ is the solid-fluid interaction force. The
contact force for particle $i$ is obtained from all contact forces
between $i$ and neighboring particles:

\begin{equation}
\mathbf{F}_{i}^{c}=\sum_{j=1}\mathbf{F}_{ij}^{c}\:,\label{eq:ContactSum}
\end{equation}
A more detailed description of the used DEM model can be found in
\cite{Oschmann2014,Markauskas2015}.

The interaction force $\mathbf{F}_{i}^{int}$ acting on solid particle
$i$ can consist from several individual solid-fluid forces \cite{Zhu2007}.
Currently the drag force $\mathbf{F}_{i}^{D}$ and the pressure gradient
force $\mathbf{F}_{i}^{\nabla p}$ are considered as the dominant
interaction forces:

\begin{equation}
\mathbf{F}_{i}^{int}=\mathbf{F}_{i}^{D}+\mathbf{F}_{i}^{\nabla p}.\label{eq:Interaction}
\end{equation}

Various models are available for the calculation of the drag force.
In the current work the correlation proposed by Di\,\,Felice \cite{DiFelice1994}
is used:

\begin{equation}
\mathbf{F}_{i}^{D}=\frac{1}{8}C_{d}\rho_{f}\pi d_{i}^{2}(\mathbf{u}_{f,i}-\mathbf{v}_{i})|\mathbf{u}_{f,i}-\mathbf{v}_{i}|\varepsilon_{i}^{2-\chi},\label{eq:DiFelice}
\end{equation}
where $\varepsilon_{i}$, $C_{d}$, $d_{i}$, $\mathbf{u}_{f,i}$,
$\mathbf{v}_{i}$ are the fluid fraction at solid particle $i$ location
calculated by Eq. \ref{eq:fluid-fraction}, the drag coefficient,
the solid particle diameter, the fluid velocity and the solid particle
velocity correspondingly.

\section{Numerical analysis of particle separation}

Wet separation of spherical particles with different densities is
analyzed numerically. As a first step for the analyses of a wet separation
problem, a set of generated particles with two different densities
is dropped into a rectangular container filled with liquid. Depending
of the density of the particles in comparison with the liquid density,
they settle down or float. This process is simulated using DEM for
the solid particles coupled with SPH for the fluid. Two simulations
of the system are performed. In the first case (denoted as S1) particles
with the densities $\rho_{1}=800\:\mathrm{kg/m^{3}}$ and $\rho_{2}=1200\:\mathrm{kg/m^{3}}$
are used. For the second case (denoted as S2), particles with the
densities $\rho_{1}=900\:\mathrm{kg/m^{3}}$ and $\rho_{2}=1100\:\mathrm{kg/m^{3}}$
are used. In both cases the 1000 monosized particles (500 with $\rho_{1}$
and 500 with $\rho_{2}$) with diameter $d=4\:\mathrm{mm}$ are generated
above the free surface of the liquid. Physical properties of water
($\rho_{f}=1000\:\mathrm{kg/m^{3}}$, $\mu=0.001$) are used for the
liquid.

\begin{figure}
\begin{centering}
a)\includegraphics[width=4cm]{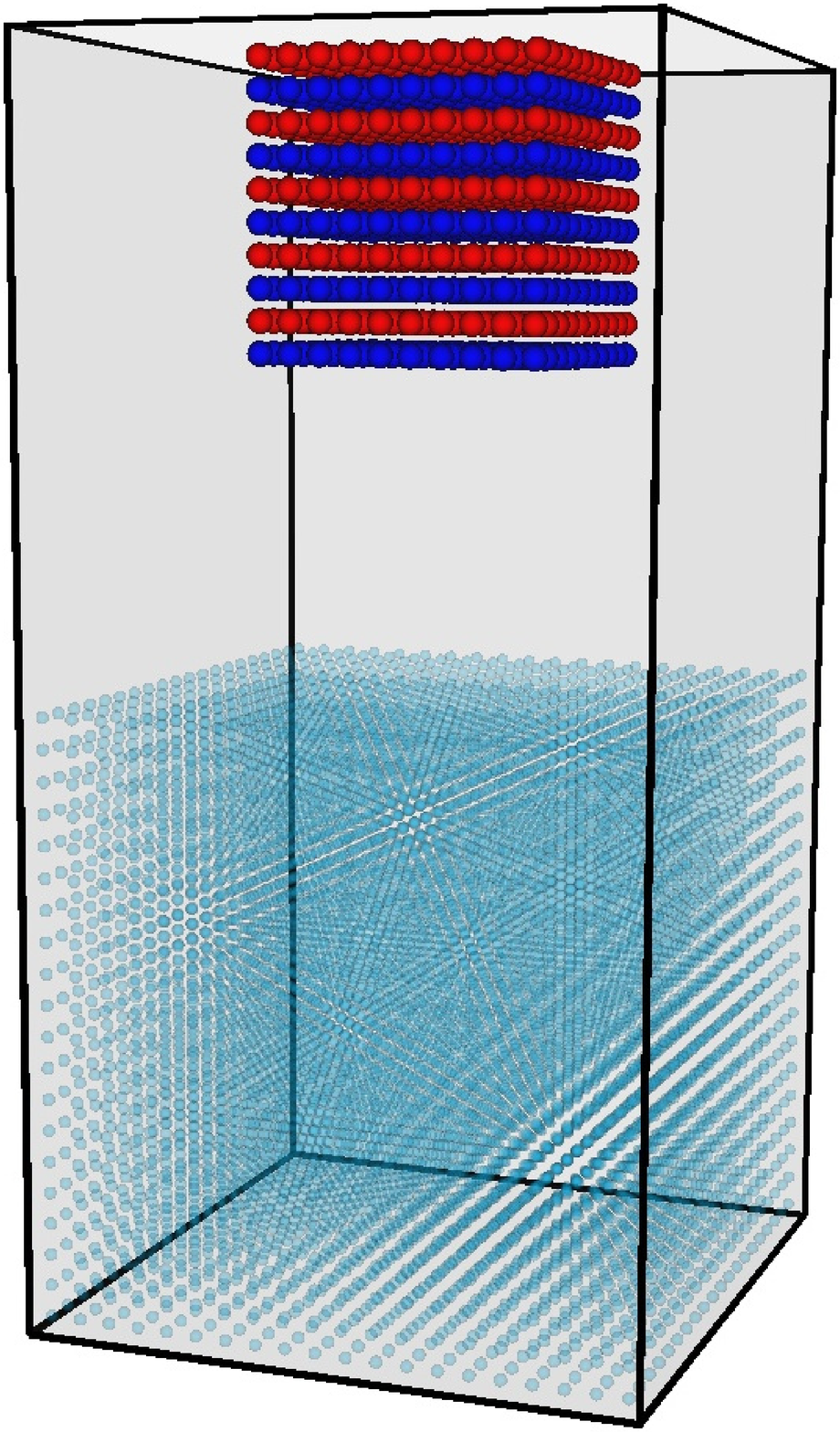} b)\includegraphics[width=4cm]{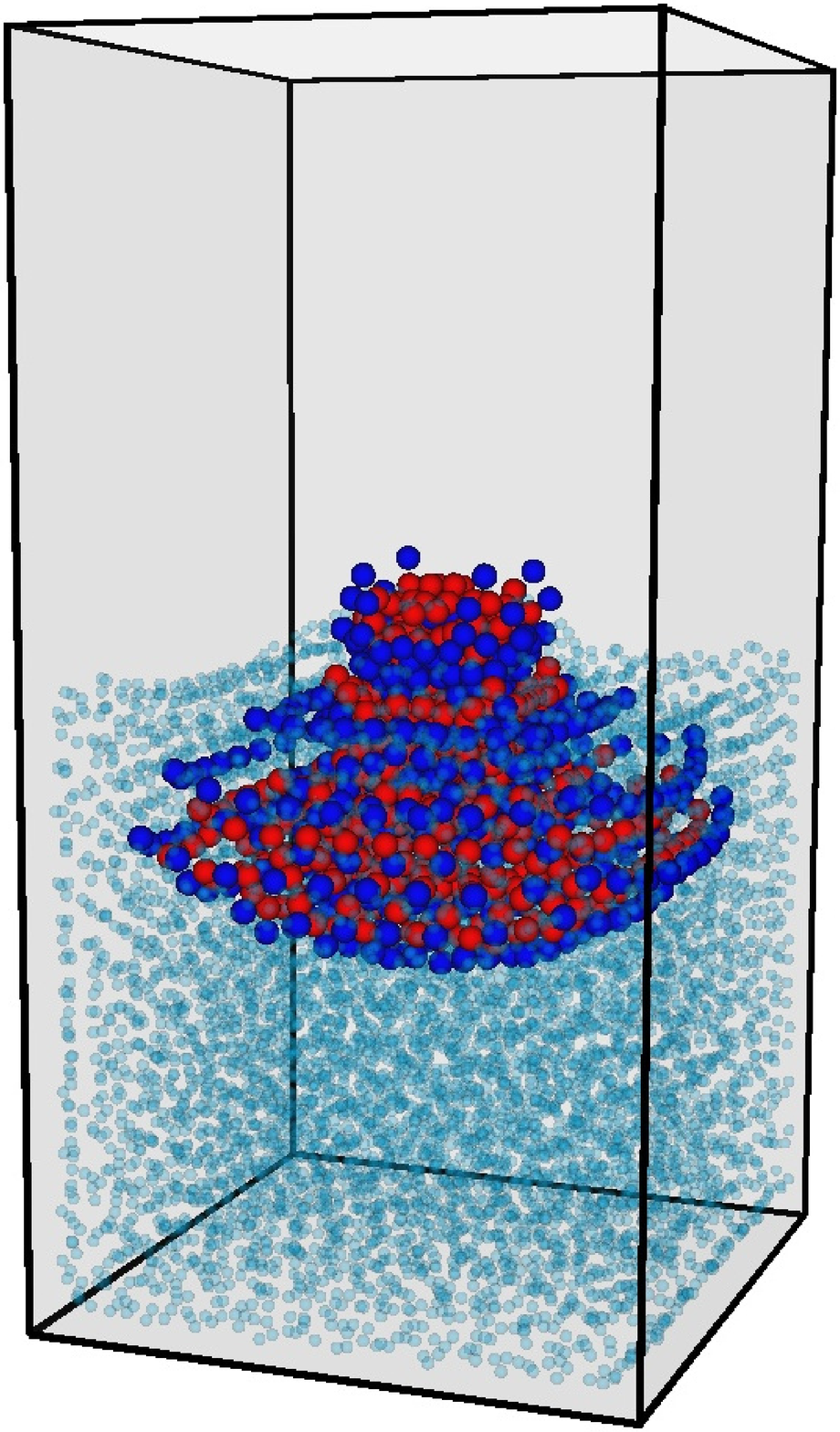}
c)\includegraphics[width=4cm]{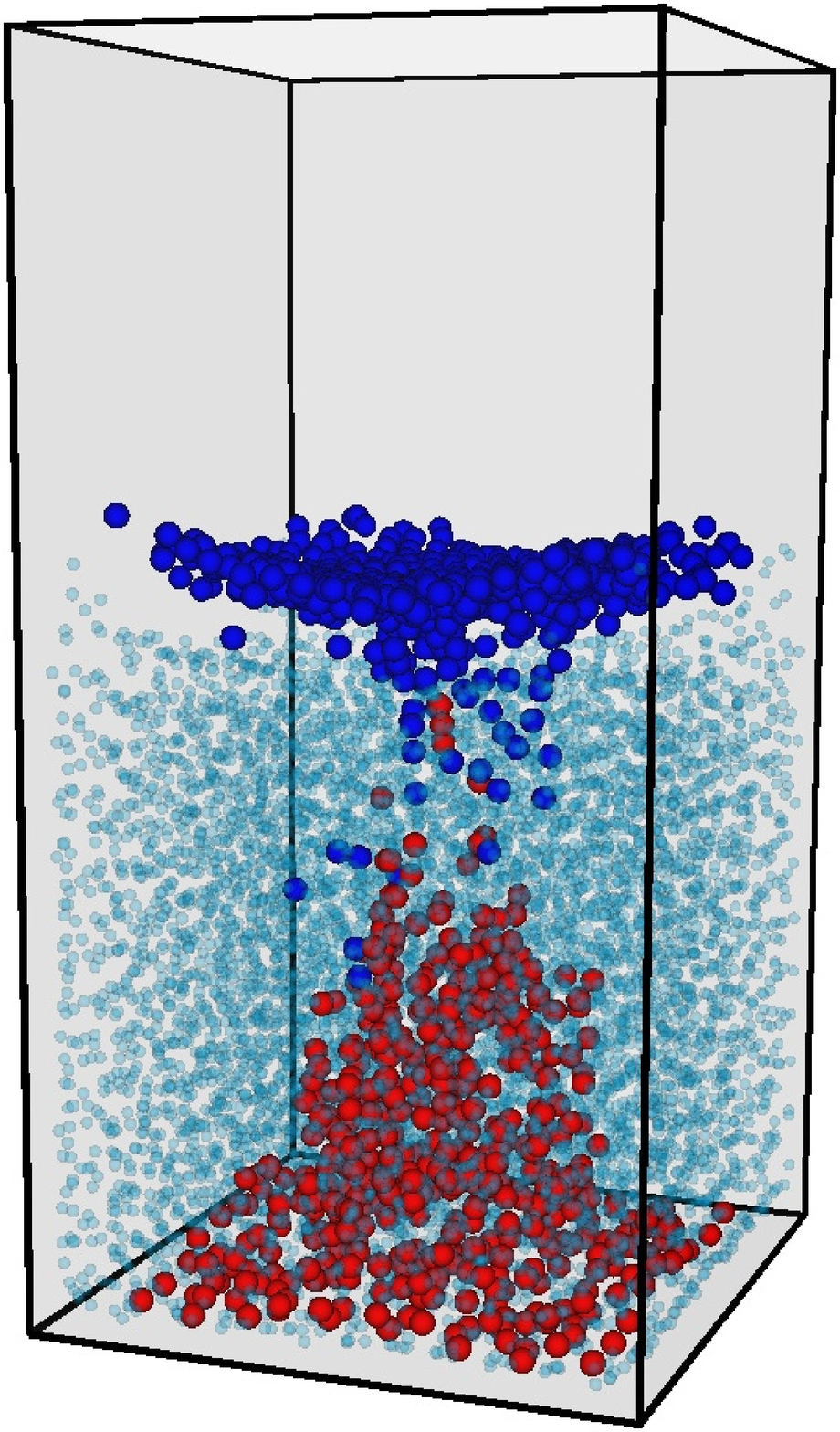}
\par\end{centering}

\caption{Particles in a rectangular container at different time instances (case
S1): a) $t=0.\,\mathrm{s}$, b) $t=0.3\,\mathrm{s}$ a) $t=1.\,\mathrm{s}$
\label{fig:Particles}}

\end{figure}

\begin{figure}
\begin{centering}
\includegraphics[width=7cm]{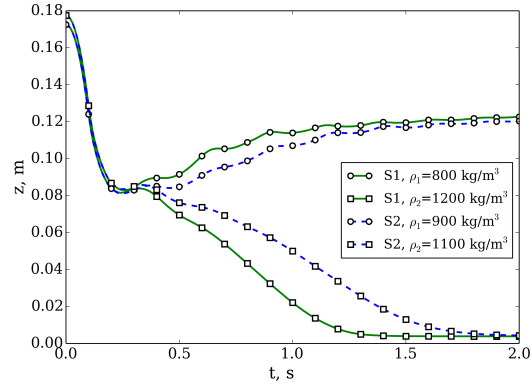}
\par\end{centering}

\caption{Evolution of averaged vertical particle coordinates depending on their
densities during the separation process \label{fig:Evolution-of-coord}}

\end{figure}

The process of particle separation is presented in Fig.~\ref{fig:Particles},
where the particle positions at different time instances are shown.
As can be seen, the particles with a higher density are settling to
the bottom of the container, while the particles with a lower density
than the density of the liquid are floating. 

The evolution of averaged vertical coordinates of the particles is
presented in Fig.~\ref{fig:Evolution-of-coord}. When the particles
are dropped, both kinds of particles initially submerged and then
start to separate. As can be expected, the particles in simulation
S2 take longer to fully separate.

\begin{figure}
\begin{centering}
\includegraphics[width=6cm]{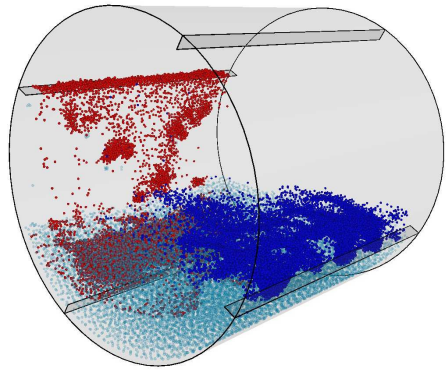}
\par\end{centering}

\caption{Wet separation of particles in a rotating cylinder \label{fig:Rotating-cyl}}

\end{figure}

The presented simulations of particle separation in the rectangular
container are used as an initial step for more sophisticated simulations
of wet particle separation in a rotating cylindrical container. The
initial result of a simulation of such a system is shown in Fig.~\ref{fig:Rotating-cyl}.
Here, the mixture of the particles is dropped into the liquid. The
sedimented particles are lifted from the bottom of the cylinder with
the help of short lifters attached to the wall and are removed later.

\section{Concluding remarks}

The presented work shows our efforts aiming to the numerical analysis
of the wet separation of particles with different densities. DEM was
used for the simulation of the solid particles, while SPH was used
for the simulation of the liquid. Some details of the used coupled
DEM-SPH method were presented. Simulations of particle separation
were performed and results were presented. It can be concluded, that
the used coupled DEM-SPH method is capable to simulate wet particle
separation. The performed simulations showed how the numerical scheme
is able to capture the dependency of the separation process on the
different particle densities.

\section*{Acknowledgements}

This project has received funding from the European Union\textquoteright{}s
Horizon 2020 research and innovation programme under the Marie Sklodowska-Curie
grant agreement No 652862.

\bibliographystyle{aipproc}
\bibliography{Markauskas_Wet-separation}

\end{document}